**Microglial memory of early life stress and inflammation: susceptibility to neurodegeneration in adulthood**

Paula Desplats [1,2], Ashley M. Gutierrez[1], Marta C. Antonelli[3,4] and Martin G. Frasch[5,6]

1. Department of Neurosciences, University of California San Diego, CA, USA
2. Department of Pathology, University of California San Diego, CA, USA
3. Instituto de Biología Celular y Neurociencia "Prof. Eduardo De Robertis", Facultad de Medicina, Universidad de Buenos Aires, Argentina
4. Department of Obstetrics and Gynecology, Klinikum rechts der Isar, Technical University of Munich, Germany
5. Department of Obstetrics and Gynecology, University of Washington, Seattle, WA, USA
6. Center on Human Development and Disability, University of Washington, Seattle, WA, USA

**Address of correspondence:**
Martin G. Frasch
Department of Obstetrics and Gynecology
University of Washington
1959 NE Pacific St
Box 356460
Seattle, WA 98195
Phone: +1-206-543-5892
Fax: +1-206-543-3915
Email: mfrasch@uw.edu

**Conflict of interest statement:** The authors have declared that no conflict of interest exists

**Abstract**

We review evidence supporting the role of early life programming in the susceptibility for adult neurodegenerative diseases while highlighting questions and proposing avenues for future research to advance our understanding of this fundamental process. The key elements of this phenomenon are chronic stress, neuroinflammation triggering microglial polarization, microglial memory and their connection to neurodegeneration. We review the mediating mechanisms which may function as early biomarkers of increased susceptibility for neurodegeneration. Can we devise novel early life modifying interventions to steer developmental trajectories to their optimum?

1. Introduction

In this focused review, we propose that susceptibility to adult neurodegenerative diseases is programmed in the womb, and early postnatal preventive measures may reduce the risk. We delineate the key mechanisms and players mediating this phenomenon, a foundation for preventive strategies. First, we discuss what are likely the most common two insults taking place during fetal development and imprinting upon fetal brain: chronic pregnancy stress and inflammation. We propose that such imprinting is mediated via the microglial plasticity resulting in memory properties and that such memory may increase susceptibility to neurodegenerative diseases in adult life. We review the animal models required for studying and testing the mechanisms. The proposed concept requires a discussion of microglial plasticity and mechanisms by which *in utero* insults influence the dynamic process of microglial-neuronal interactions in the developing brain. The role of microglia in adult neurodegenerative disorders is discussed and this concept is brought back to the perinatal origins of adult susceptibility to neurodegeneration. Finally, we discuss how our present understanding of the mechanisms of microglial memory formation may inform novel modifying treatments to steer developmental trajectories to their optimum.

2. Programming of fetal brain development by chronic stress exposure: the concept

The complex development of the nervous system is characterized by essential processes like cell proliferation, migration and differentiation that follow a tightly regulated program at well-coordinated time points to ensure the establishment of brain structures and functions (Andersen 2003). During this developmental process, the organisms may be exposed to insults and environmental influences that might be critical for later susceptibility to diseases. Several models have been proposed to explain how this exposure to insults *in utero* and/or during early postnatal stages affects the development of target organs, disrupts homeostasis and increases the risk for diseases manifested in later life. Based on his studies on adult cardiovascular disorders, Barker proposed the hypothesis of the Fetal Basis of Adult Diseases (FeBAD) (Barker 1993) . FeBAD suggests that the fetus responds to the maternal health status mounting adaptive responses for survival. Stemming from Barker´s hypothesis, several other models have been proposed (Gluckman and Hanson 2006, Ben-Ari 2008, Gluckman, Hanson et al. 2010) prompting Van den Bergh to formulate the hypothesis of the "Developmental Origins of Behavior, Health and Disease" (DOBHaD), which integrates early brain and behavioral development with new insights from the field of epigenetics (Van den Bergh 2011).

Taken together, these models point to the concept that adult health and behavior are programmed *in utero*; and they are shaped postnatally throughout lifetime through new exposures. This continuous programming is mediated by epigenetic mechanisms to which we will return in Section 6.

Perinatal stress is a known powerful driver of an individual's spatiotemporal epigenome and integrative physiological responses: different cell types and organs - particularly the brain - elicit specific stress responses at different developmental stages due to genomic-/epigenomic-environmental interactions (Metz, Ng et al. 2015, Conradt, Adkins et al. 2018, Frasch, Lobmaier et al. 2018). On the integrative physiological scale in humans, chronic maternal stress in the early third trimester of gestation can program heart rate regulatory systems in the fetus to be entrained by maternal heart rate fluctuations. This phenomenon is not seen in control subjects with low levels of stress as measured by Cohen perceived stressed scale (PSS-10) (Lobmaier S. 2019). More generally, such relative coordination and entrainment phenomena are well-described in the weakly coupled nonlinear dynamical systems, such as the complex physiological systems (Scott Kelso, Holt et al. 1981, Hoyer, Frasch et al. 2001).

Despite these pilot reports in human cohorts, much of the evidence on the effects of perinatal stress on adult life has been gathered from animal studies (Frasch, Baier et al. 2018). We believe that animal models will continue to be indispensable to investigate the alterations of glial-neuronal development under various insults, including perinatal stress, to understand their mechanistic contribution to increased susceptibility to adult neurodegenerative diseases.

## 3. Animal models of perinatal stress reveal links to early life neuroinflammation and neurodegeneration

The last twenty years have witnessed an exponential increase in publications related to the effect of perinatal stress on offspring neurodevelopment both in animal models and in humans (Jawahar, Murgatroyd et al. 2015, Van den Bergh, van den Heuvel et al. 2017, Frasch, Baier et al. 2018). A detailed description of the factors that might lead to perinatal stress has been summarized before (Frasch M 2017). Broadly, these factors are comprised of a vast number of modifiable and nonmodifiable influences that the mother-child dyad might experience pre- and/or postnatally. The modifiable factors include under and overnutrition, substance abuse, drugs intake, intrauterine growth restriction or infection. The nonmodifiable factors include wars, famines, and natural disasters. Perinatal stress includes both, models of *pre*natal stress (PS) (Hubacek, Pitha et al. 2000) and *post*natal stress; the latter are often referred to as Early Life Stress (ELS) (Jawahar et al, 2015). Note that in this review we reserve the PS abbreviation to *pre*natal stress.

Animal models of PS have been designed to transfer the stress experienced by the mother to the embryo or fetus *in utero* by exposing the pregnant dam to stressors during different gestational phases (Huizink, Mulder et al. 2004, Mastorci, Vicentini et al. 2009, Bock, Wainstock et al. 2015). Similarly, animal models of ELS have been designed to mimic childhood adversities, mostly related to the quality of maternal care such as maternal neglect or maltreatment (Jawahar, Murgatroyd et al. 2015, McDonnell-Dowling and Miczek 2018).

Regardless of the type of stressor applied, its frequency, duration and gestational/postnatal period of application, most studies using animal models provided strong evidence that exposure to insults during the prenatal period or during early postnatal development turns the offspring vulnerable to disease in later life. The consequences of such stressors are broad and can include learning and cognitive delays, cardio-metabolic alterations, pro-inflammatory responses, neuroendocrine impairment and neurodegenerative disorders (Van den Bergh 2011, Modgil, Lahiri et al. 2014, Heim, Entringer et al. 2019). An important example of this relationship is observed in the limited nesting and bedding model (Rice et al, 2008; Walker et al, 2017), a specific paradigm of ELS related to maternal neglect and maltreatment which was shown to exacerbate the amyloid-$\beta$ (A$\beta$) plaque burden in the hippocampus and the neuroinflammatory response to A$\beta$ in aged transgenic APP/PS1 mice (Brunson, Kramar et al. 2005, Hoeijmakers, Ruigrok et al. 2017). Conversely, enhanced maternal care after postnatal handling reduces amyloid accumulation in the hippocampus ameliorating cognitive decline in the APPswe/PS1dE9 mouse model of Alzheimer's disease (AD) (Lesuis, van Hoek et al. 2017).

Together, these observations led to the neurodevelopmental hypothesis of neurodegenerative disorders: an unfavorable environmental input during early life shapes adult neuroplasticity that might result in neurodegenerative processes (Schaefers and Teuchert-Noodt 2016, Fanni, Gerosa et al. 2018).

This neurodevelopmental hypothesis is clearly related to the "two hits" hypothesis in which each of the environmental inputs is considered a "hit" and, after a certain latency period, a second trigger is necessary for the disease to develop (Lahiri, Maloney et al. 2009). Nowadays, the theory has gained consideration as the concept of "multiple hits" to better interpret the composite genetic and environmental interplay that contributes to the development of neurodegenerative diseases (Faa, Marcialis et al. 2014, Davis, Eyre et al. 2016, Post, Lieberman et al. 2018).

For example, prenatally stressed ("first hit") adult rats exposed to an intrastriatal injection of 6-hydroxy-DA (6-OHDA) to damage dopaminergic neurons ("second hit") show high levels of tyrosine hydroxylase (TH) positive cells in the ventral tegmental area (VTA) which are more susceptible to 6-OHDA. In addition, the number and asymmetry of neuronal nitric oxide synthase (nNOS)-expressing cells in the VTA and nucleus accumbens are altered in PS rats. These results suggest a deregulation between dopaminergic and nitrergic systems in the VTA of PS rats, which responded to the neurotoxic aggression by increasing the number of nNOS positive cells involved in neuroprotection of TH positive cells (Baier et al., 2014).

PS affects both neuronal and glial cells, which show severe alterations in morphology and neurotransmitter systems. Adult male offspring of stressed rat dams exhibits long-lasting astrogliosis with reduced dendritic arborization (Barros, Duhalde-Vega et al. 2006). This is accompanied by an increase of ionotropic and metabotropic glutamate receptors in the frontal cortex (FCx), striatum and hippocampus (Berger, Barros et al. 2002). In addition, there is an elevation in vesicular vGluT1-transporter in the FCx and GLT1 in hippocampus, associated with an increased glutamate uptake capacity in the FCx (Adrover, Pallares et al. 2015) and alterations in nitric oxide synthase-expressing cells (Baier, Pallares et al. 2014).

Furthermore, PS induces changes in the glutamatergic metabolism, such as reduction of glutamatergic neurotransmission in the ventral hippocampus (Marrocco, Mairesse et al. 2012) and sex-specific increase of glutamate-receptor expression (Zuena, Mairesse et al. 2008, Wang, Ma et al. 2016). In addition, enhanced microglial activation is evidenced by an increased expression of microglial markers (Slusarczyk, Trojan et al. 2015), increased number of microglial cells with large somas (Diz-Chaves, Pernia et al. 2012), and an acceleration of microglial differentiation into ramified forms in the pre-weaning offspring (Gomez-Gonzalez and Escobar 2010).

In sum, piling evidence derived from the PS animal models shows significant reprogramming of the glutamatergic-nitric oxide-microglia metabolism in prenatally stressed offspring. This supports the notion that exposure to insults during the prenatal and/or early life period may shape the vulnerability to neurodegenerative disorders in later life.

4. **Microglial plasticity due to fetal and neonatal insults: a microglial memory programming mechanism for life**

Microglia are the resident macrophages of the central nervous system, involved in immune surveillance; synaptic pruning and phagocytosis of misfolded protein aggregates (Heppner, Ransohoff et al. 2015).

The neuropsychological notion of memory is usually linked to the concept of neuronal plasticity, i.e., the ability of neuronal cells and neuronal ensembles to adapt to their environment, individually and collectively, and respond adaptively when re-encountering the same or similar set of stimuli. Evidence is mounting that similar mechanisms operate in glial cells, particularly microglia. Beyond their role as the "conventional macrophages of the brain," microglia have fundamental roles in synaptogenesis (Kettenmann, Hanisch et al. 2011). As we discuss in detail in the next section, aberrant glial-neuronal interactions are important players in early stages of AD and other adult neurodegenerative diseases (Madhusudan, Vogel et al. 2013, Bilbo and Stevens 2017, Frasch, Burns et al. 2018).

It is well known that excessive activation of ionotropic glutamate receptors causes excitotoxicity and promotes cell death, underlying a potential mechanism of neurodegeneration. In addition to the excitotoxicity and inflammatory processes that characterize neurodegenerative disorders, an emergent role has recently been discussed for the nitric oxide pathways in relation to the glutamate neurotoxicity (Manucha 2017). Glutamate receptors are also expressed in microglia and it has been postulated that a dysfunctional subunit GluA2 may accelerate glutamate neurotoxicity via release of proinflammatory cytokines from microglia in neurodegenerative diseases (Noda 2016).

Overactivation of microglia with a memory of PS may inflict injury in adulthood. For example, PS paradigms such as models of depression affect the offspring later in life inducing behavioral changes in adult mice, elevated release of chemokines and a profuse microglial activation (Slusarczyk, Trojan et al. 2015). In addition, PS has been linked with an abnormal expression of the α7 nicotinic acetylcholine receptor (α7nAChR) in the male PS-exposed adult brain (Baier, Pallares et al. 2015). Prenatal exposure to maternal restraint stress, a form of PS, reduced protein levels of α7nAChR in the adult frontal cortex and hippocampus, key areas affected by the AD pathology (Jellinger and Attems 2013). Further studies are needed to associate cell type-specific expression of α7nAChR to PS as well to neuropathological and behavioral phenotypes in mid and late adulthood.

Recent studies in mice also indicate the role of inflammation in shaping microglia and behavior in the aging offspring of stressed mothers. Early life exposure to lipopolysaccharides (LPS), a widely used model of inflammation caused by infection with gram-negative bacteria, results in differential location of activated microglia in the brain, with sustained alterations in hippocampal microglia that become hypersensitive to inflammatory markers. These changes endure throughout adulthood and have negative effects on memory and learning (Schaafsma, Basterra et al. 2017).

In line with these LPS-based studies and independent of PS, infections of the gut may precede Parkinson's disease (PD) by years or trigger it in an experimental mouse model (Matheoud, Cannon et al. 2019, Nerius, Doblhammer et al. 2019). The latter demonstrated involvement of glial cells in the pathogenesis of the autoimmune response resulting in a PD phenotype. These findings underscore the importance of the brain-gut axis for neurodegenerative diseases and the likely multi-hit nature of their etiology. The longest brain nerve connecting brain and body is the vagus nerve, an important substrate of cholinergic signaling, including peripheral signaling via the α7nAChR, the cholinergic anti-inflammatory pathway (Rosas-Ballina, Olofsson et al. 2011). This brain-body connection begins *in utero* (Garzoni, Faure et al. 2013) and interruption of the vagus nerve signaling reduces the risk for PD (Liu, Fang et al. 2017).

Non-neuronal (i.e., non-vagus-nerve), humoral, lymphatic and blood-stream - mediated brain-gut pathways have also been discussed in linking enteral and cerebral homeostasis (Santos, de Oliveira et al. 2019).

In addition to gut infections, chronic oral cavity infections have also been associated with an increased risk for AD, probably mediated by resident microglia responding to chronic systemic inflammation caused by periodontitis (Kamer, Craig et al. 2008, Wu and Nakanishi 2014).

Are there common features of the reported studies that could be integrated into a microglial endophenotype that increases the risk for neurodegeneration in adulthood? Perhaps an immunosenescence perspective can help shed light on possible mechanisms of early-life programming of neurodegeneration (Fulop, Larbi et al. 2017). What distinguishes the centenarians that remain untouched from neurodegenerative conditions? The common denominator appears to be their subjective innate or acquired resilience (Krohne 2001, Franceschi and Bonafe 2003, Southwick S 2012). Is the subjective processing of uncertainty as "less or not stressful" a key to long life with reduced immunosenescence and hence delayed or no neurodegeneration (Peters, McEwen et al. 2017)?

Peters et al. propose that there is a cerebral metabolic cost to processing uncertainty in information, a hallmark of chronic stress (Peters, McEwen et al. 2017). Early life inflammatory exposures reprogram fetal microglia toward a distinct immunometabolic phenotype defined by what teleologically appears to be energy preservation (Cao, Cortes et al. 2015). Taken together, does this implies that stress/uncertainty-imposed energy demands of information processing cannot be met fully by a stressed brain? Indeed, microglial senescence is discussed as a contributor to age-related neurodegeneration (Luo, Ding et al. 2010, von Bernhardi, Eugenin-von Bernhardi et al. 2015). Little is known about the control mechanisms governing the metabolic changes within the microglia which occur to meet the requirements of their phenotypic switch (Borst, Schwabenland et al. 2018). Similar to our findings in ovine fetal microglia in response to inflammation, metabolic changes do correlate with the ability of innate immune cells to show hallmarks of adaptive immune responses (Borst, Schwabenland et al. 2018).

Is this immunometabolic endophenotype inherited multigenerationally via the DoHAD pathway of immunosenescence (Hanson, Cooper et al. 2016)? Fetal microglia and astrocytes exhibit memory of preceding insults (Cao, Cortes et al. 2015, Frasch 2018, Cao, MacDonald et al. 2019). Does such plasticity and memory of experienced insults come at a cost of a reduced energy reserve for future hits? Does such reduced reserve translate into an increased susceptibility to adult neurodegenerative diseases (Paolicelli, Bolasco et al. 2011)?

In summary, increasing support for mechanisms representing microglial memory emerge, as well as new questions as to whether these new players in the etiology of neurodegenerative diseases also lend themselves as novel therapeutic targets which modulate the microglial memory by interfering with its programming.

## 5. Role of Microglia in Adult Neurodegenerative disorders: early life origins

Inflammation is a prominent axis of neuropathology common to most neurodegenerative disorders in adulthood. Stress has also been coming increasingly into focus as an important contributor, perhaps even as a trigger of clinical manifestation. The involvement of exacerbated microglia and astroglia reactivity in the pathophysiology of neurodegeneration is indisputable, but their precise roles and patterns of interaction with triggers such as inflammation or stress remain subject of ongoing research.

*Alzheimer's disease, neuroinflammation and stress*

AD is a multifactorial neurodegenerative disorder that affects nearly 50 million people worldwide, representing the most prevalent form of dementia in older adults. AD is manifested clinically by progressive memory loss and a gradual decline in global cognitive function that interferes with daily living. At the neuropathological level, AD is characterized by accumulation of amyloid-$\beta$ (A$\beta$) in amyloid plaques, phosphorylated Tau protein in neurofibrillary tangles, and progressive neuronal loss in the neocortex and the hippocampus (Querfurth and LaFerla 2010).

Inflammation is a major feature of AD pathology (Gorelick 2010). Cytokines, chemokines and complement factors are found in the cerebrospinal fluid (CSF) and amyloid plaques in the brain of AD patients. Several epidemiological studies and clinical trials link inflammatory markers to dementia and cognitive impairment (Beard, Waring et al. 1998, Aisen, Schafer et al. 2003, Barber 2011). Moreover, the expression of inflammation-related genes varies as disease progresses, suggesting their role in pathophysiology (Bossers, Wirz et al. 2010).

Ageing and neurodegeneration diminish microglial function, as evidenced by disruption of microglia-specific transcripts such as TREM2 and CD33 in AD brains (Clayton, Van Enoo et al. 2017). Interaction of microglia with A$\beta$ plaques activates the pro-phagocytic phenotype (M2) to clear the aggregates. However, as the plaque burden increases in the brain overtime, microglia shift towards a pro-inflammatory phenotype (M1), triggering chronic inflammation and neuronal death (Tang and Le 2016). A$\beta$-stimulated microglia show high levels of phagocytic markers such as *AXL*, *CLEC7A*, *LGALS3*, *CD11C*, *TREM2*, and *TYROBP.* While this activation may be beneficial for the clearance of protein aggregates, a hyperactive response to A$\beta$ has been reported in the brains of early onset AD cases and may be associated with neuronal death (Yin, Raj et al. 2017).

While amyloid plaques are formed by the assemblage of large insoluble fibrils, smaller soluble Aβ oligomers can spread in the brain and are the most toxic forms that induce synaptic dysfunction (Smith and Strittmatter 2017). Interaction of these oligomers with TLR2 receptors in microglia induces a neurotoxic response with increased transcription of pro-inflammatory molecules like iNOS and TNFα (Heurtaux, Michelucci et al. 2010, Liu, Liu et al. 2012). Furthermore, Aβ oligomers persistently activate the tyrosine kinase pathway sustaining the release of neurotoxic factors (Dhawan, Floden et al. 2012) and the upregulation of TNFα and cytokines IL-1 and IL-6 (Meda, Cassatella et al. 1995).

Remarkably, studies in genetic mouse models of AD and indirect results in non-human primates indicate that chronic stress increases Aβ-mediated neuronal toxicity or increases Tau accumulation or phosphorylation and tangle formation (Dong and Csernansky 2009). A sustained neuroinflammation was suggested as a link between the initial Aβ pathology and the later tangle formation (Kinney, Bemiller et al. 2018). Future animal studies need to examine whether removing the microglial memory of prenatal insults via PS or infection may halt or at least slow down the progression of the three core AD features, Aβ, tangles and the chronic inflammation and, ultimately, the behavioral dementia phenotype.

*Parkinson's disease, neuroinflammation and stress*

PD is the second most frequent neurodegenerative disorder of the elderly, characterized clinically by motor impairment, gait disturbance, tremor and increased risk to dementia. At the neuropathological level, PD brains present with an accumulation of of α-synuclein (α-syn) aggregates in Lewy body structures, mainly in the substantia nigra (SN) and loss of dopaminergic neurons (Spillantini, Schmidt et al. 1997, Takeda, Mallory et al. 1998, Burn 2006). PD is a multifactorial disorder where genetic and environmental factors converge to generate pathology, and more than 90% of cases are of sporadic origin. Emerging data supports that chronic neuroinflammation is highly prevalent in PD, and increased number of microglia and microgliosis have been reported in postmortem PD brains and transgenic animal models of PD (Watson, Richter et al. 2012, Doorn, Moors et al. 2014). Importantly, microgliosis is more evident in the vicinity of dopaminergic neurons primed for degeneration in the SN, hippocampus and cortex (McGeer, Itagaki et al. 1988, Banati, Daniel et al. 1998, Imamura, Hishikawa et al. 2003, Sawada, Imamura et al. 2006). Activated microglia produce high levels of pro-inflammatory molecules like TNFα, IL6, nitric oxide synthase (NOS) and cyclooxygenase-2 (COX-2) that contribute to the cell-specific vulnerability in PD (Hunot, Boissiere et al. 1996, Knott, Stern et al. 2000, Block, Zecca et al. 2007, Dufek, Rektorova et al. 2015). Indeed, a higher density of activated microglia was observed in nigral neurons of mice exposed to the pro-inflammatory bacterial LPS. This region-specific response may render the SN more vulnerable to neuronal death than cells in the cortex or hippocampus correlating with the disease pathology (Kim, Mohney et al. 2000).

Exacerbated neuronal death, on the other hand, can also trigger microglial activation thus feeding a deleterious loop. Substances which are produced by dying dopaminergic neurons that can activate microglia include α-syn aggregates, ATP, matrix metalloproteinase-3 (MMP-3) and neuromelanin (Wilms, Rosenstiel et al. 2003, Davalos, Grutzendler et al. 2005, Zhang, Wang et al. 2005, Kim, Choi et al. 2007). Notably, suppression of microglial activation alleviates the PD phenotype in mouse models, further supporting the role of these cells in the pathophysiology (Wu, Jackson-Lewis et al. 2002).

Last but not least, chronic stress, including ELS, has been reported to contribute to PD etiology, potentially triggering PD (Hemmerle, Herman et al. 2012, Djamshidian and Lees 2014, Dalle and Mabandla 2018).

*Innate immune system, ELS, inflammation and neurodegeneration in adult life*

The role of the innate immune system in AD has gained attention after several genome-wide association studies identified disease-associated genomic risk loci that map to immune response genes and genes expressed by microglia. A recent study identified 29 risk loci associated to 215 genes with potential of causing AD; many of them were highly expressed in immune-related cells, including microglia (Jansen, Savage et al. 2019). Among these, *INPP5D; GPSM3; PILRA; MS4A6A; MS4A7; RIN3* and *HMHA1* are microglia-enriched genes expressed in brain regions relevant to AD and associated with Braak staging and as measure of disease progression (Kunkle, Grenier-Boley et al. 2019). Interestingly, two studies identified a single-nucleotide polymorphism (SNP) at the third intron in *KAT8* as a new AD risk locus. Multiple variants in this locus influence *KAT8* expression and methylation levels, thus also linking epigenetic changes to AD risk (Marioni, Harris et al. 2018, Jansen, Savage et al. 2019).

While evidence supports the role of neuroinflammation in disease causation, the mechanisms that trigger this process and how inflammation reaches the brain are still not fully understood. One plausible explanation is that AD involves systemic alterations rather than being restricted to the brain. The innate immune response system eliminates pathogens in a fast and non-specific reaction involving inflammation. In the periphery, this system is composed by neutrophils, monocytes/macrophages, dendritic cells and natural killer cells, while in the brain this function is fulfilled by the resident microglia (Guillot-Sestier, Doty et al. 2015, Crotti and Ransohoff 2016, Le Page, Dupuis et al. 2018). Adaptive immunity, however, cannot be initiated directly in the brain. Once microglia are activated in response to an insult, secretion of cytokines and chemokines is increased, attracting circulating lymphoid and myeloid cells and thus establishing a communication between the brain and the systemic immune response mechanisms (Ransohoff and Brown 2012). In addition, alterations in the brain blood barrier observed in most neurodegenerative diseases, facilitate the flow of inflammation-related molecules between the brain and the systemic circulation, enabling the cross-talk between these systems (Busse, Michler et al. 2017). Therefore, adverse events during early life may affect the communication between brain and periphery paving the road for inflammatory-initiated diseases during adulthood.

Prenatal challenging of the immune system by exposure of mice to Poly I:C (polyriboinosinic–polyribocytidilic acid) during gestation as a model of viral infections, results in chronic elevation of inflammatory cytokines, higher levels of amyloid precursor protein (APP) in the hippocampus, increased Tau phosphorylation and defective working memory in old age animals - features directly associated with AD pathology. These alterations are accompanied by microglial activation and reactive gliosis suggesting that priming the microglial system during early life favors a chronic inflammatory state that seeds neurodegenerative changes in the adult (Krstic, Madhusudan et al. 2012). As emerging evidence links viral infections with AD pathology (Readhead, Haure-Mirande et al. 2018), a provocative question is whether early-life exposure to viruses may contribute to this increased risk for AD.

Multiple environmental toxins, particularly agrichemicals, have been linked to increased risk for PD (Wang, Costello et al. 2011, Desplats, Patel et al. 2012) and in many cases inflammation seems to be the underlying mechanism. Emerging evidence suggests that infections during the prenatal life also impact PD susceptibility (Collins, Toulouse et al. 2012). Maternal infections like bacterial vaginosis may increase the risk of the neonate to develop PD, an effect apparently mediated by the rise in chorioamniotic LPS and TNF-α levels in infected pregnant women, that hinders the development of dopaminergic neurons (Okada, Matsuzaki et al. 1997, Ling, Gayle et al. 2002).

Similarly, prenatal exposure to LPS increases the vulnerability of rats to rotenone in later life, resulting in synergistic effects that increase loss of tyrosine-hydroxylase (TH) cells in the SN in comparison to LPS-naïve animals. Dopaminergic cell loss was associated with increased levels of TNF-α and reactive microglia (Ling, Chang et al. 2004). Other studies also support the idea that viral infections in the prenatal environment can prime the organism to develop PD as adults, potentially due to changes in density and/or activation states of microglia. *In utero* exposure to the synthetic dsRNA analogue Poly I:C increases the production of pro-inflammatory cytokines in the brain of exposed rats (Cunningham, Campion et al. 2007). Importantly, these events aggravate neurodegeneration inducing severe decay of the nigrostriatal dopaminergic system in adult animals (Deleidi, Hallett et al. 2010, Field, Campion et al. 2010). While the specific role of microglia in mediating prenatal infections, dopaminergic neuronal loss and PD risk is yet to be defined, taken together, these studies point at inflammation as an important agent contributing to PD risk via a modulation of the microglial phenotype.

## 6. Epigenetic mechanisms link prenatal adverse environment, microglial memory and activation with adult neurodegenerative disorders

Epigenetic mechanisms that dictate chromatin accessibility, such as DNA methylation and post-transcriptional histone modifications, modulate gene expression and are dynamically regulated in post-mitotic neurons. As these mechanisms have crucial functions in memory formation and synaptic plasticity (Day and Sweatt 2010, Sen 2015, Cholewa-Waclaw, Bird et al. 2016), it is not surprising that alterations in epigenetic regulation play a role in neurodegeneration. For example, AD brains show a general decay in global methylation and acceleration of aging as measured by methylation markers, particularly in areas selectively vulnerable to pathology (Urdinguio, Sanchez-Mut et al. 2009, Mastroeni, Grover et al. 2010, Levine, Lu et al. 2015). In PD brains, we observed aberrant cytoplasmic localization of DNA methyl transferase 1 (DNMT1) in neurons due to interaction with aggregated α-syn resulting in reduced global methylation, including genes associated with PD pathology (Desplats, Spencer et al. 2011).

*Epigenetic modulation of microglial memory due to inflammation*
Notably, epigenome-wide association studies in cognitively impaired subjects and AD cases identified significant changes in DNA methylation at 71 methylation sites across the genome, that correlated with AD pathology burden. Among the differentially methylated genes were *ANK1* and *RHBDF2* which are involved in the signaling cascade that modulates the activation of microglia and the infiltrating macrophages to release TNFα from the cell surface (De Jager, Srivastava et al. 2014). More recently, *ANK1* was reported to be specifically upregulated in microglia isolated from the hippocampus of AD brains, without significant changes detected in neurons or astrocytes from the same samples (Mastroeni, Sekar et al. 2018). Therefore, methylation changes associated with AD pathology appear to regulate the activation of microglia and to play important roles in neuroinflammation.

Interestingly, work using inducible AD mouse models showed that early stages of pathology are associated with dynamic changes in chromatin structure nearby immune response genes. These changes include increased accessibility for the ETS family of transcriptional regulators, like PU.1, which are implicated in activation of macrophages. When mapped to orthologous regions in the human genome, the enhancers activated in the mice hippocampus corresponded to immune cell expression quantitative trait loci, further implicating immune processes in AD predisposition, and the involvement of chromatin remodeling in triggering inflammatory responses (Gjoneska, Pfenning et al. 2015).

As mentioned before, another microglial factor, TREM2, has gained attention in AD research following the identification of genetic variants in this gene associated with increased disease risk (Jin, Benitez et al. 2014). TREM2 is expressed by activated microglia and presents with differential DNA methylation in AD brains (Smith, Smith et al. 2016) supporting a role for the epigenetic machinery in the modulation of the microglial transcriptome that may cause neurodegeneration.

Key genes at almost every level of the inflammation response appear to be regulated by DNA methylation, including IL-2, CXCL2 and chemokine receptors CXCR4 and CCR7 (Bruniquel and Schwartz 2003, Mori, Kim et al. 2005, Murayama, Sakura et al. 2006, Ramos, Grochoski et al. 2011, Fischer, Onur et al. 2012). Moreover, the regulation of IL-6 requires the cross-talk between different epigenetic players including TET2 to regulate DNA methylation and HDAC2 to modulate histone acetylation, a process that reduces inflammation by repression of IL-6 in macrophages (Zhang, Zhao et al. 2015).

Microglia activation may also be dependent on histone modifications. Suh *et al.* showed that treatment with histone deacetylase inhibitors (HDACi) decreases the release of cytokines, including TNFα, IL-10, IL-1α and IL-6 suggesting that histone modifications may repress the microglial immune responses (Suh, Choi et al. 2010). More recently, using a mouse model where knock down of *Hdac1* and *Hdac2* was specifically directed to microglia, Datta et al. demonstrated that while these enzymes are not required for the survival of adult microglia in the normal brain, the deletion of Hdac1 and Hdac2 in the context of AD pathology increases the phagocytic activity of microglia, reducing amyloid plaques and cognitive decline (Datta, Staszewski et al. 2018).

Datta et al. also investigated the effects of prenatal ablation of HDACs, taking advantage of their inducible model. Interestingly, the lack of these histone-modifying enzymes early in life alters microglial proliferation and increases apoptosis, likely due to activation of cell death-signaling pathways at the transcriptional level (Datta, Staszewski et al. 2018). In a similar context, prenatal exposure to HDACi valproic acid results in higher density of microglia as well as modification of histones, further suggesting that modulation of histone changes directly affects microglia abundance (Kazlauskas, Campolongo et al. 2016). This epigenetic cascade seems to propagate across CNS cell types: activated microglia induce histone deacetylation in astrocytes reducing glutathione (GSH) transcription (Correa, Mallard et al. 2011). Low levels of GSH, in turn, trigger inflammatory molecules such as TNF-α, IL-6 and also stimulate kinase pathways further exacerbating the inflammatory state in neurodegenerative disorders (Lee, Cho et al. 2010, Pocernich and Butterfield 2012).

A more recent study supports the notion that immune memory in macrophages is executed by epigenetic alterations. Changes in histone methylation (H3K4me1) and acetylation (H3K27ac) associate with activation of enhancers that modulate inflammation-related genes. Notably, this study demonstrated that peripheral inflammation induces epigenetic remodeling in brain-resident microglia which then become tolerant to further inflammatory stimuli for at least 6-months, a phenomenon that alleviates amyloidosis in a mouse model of AD (Wendeln, Degenhardt et al. 2018). These results support the role of ELS-induced microglial memory in shaping the trajectory of AD pathology in later life.

*Epigenetic modulation of microglial memory due to psychological and nutritional stress*

Epigenetic mechanisms also mediate the effects of maternal behavior modulating microglial responses in the offspring's brain. Increased maternal care stimulates the expression of IL-10 in newborn rats in a model of drug addiction. The effect, mediated by a reduction in IL-10-promoter methylation, appears to be specific to microglia. Elevated IL-10 is sustained throughout adulthood in the Nucleus accumbens and reduces microglial reactivity after drug exposure preventing reinstatement of morphine addiction in adult animals (Schwarz, Hutchinson et al. 2011). While this study was focused on a different paradigm, it provides another example on how epigenetic remodeling is a likely mediator of early life experience that can modulate microglial responses and disease risk in later life. Future studies in animal models and human cohorts may identify epigenetic biomarkers of ELS that predict increased risk to neurodegenerative diseases in adult life.

Chronic stress may impact negatively the dietary choices (Yau and Potenza 2013), so the impact of diet on neurodegeneration is important to consider in the present context. Diet also has crucial roles in shaping the early life environment, and the consequences of malnutrition or dietary deficiency may be long-lasting. For example, maternal immune activation due to exposure to inflammatory stimuli induces persistent changes in the fatty acid composition of the fetal brain. These alterations in neurodevelopment are associated with a reduction in the levels of n-3 polyunsaturated fatty acids (e.g., omega-3) in the offspring that are manifested later in life as memory and behavioral impairments (Labrousse, Leyrolle et al. 2018, Yam, Schipper et al. 2019).

Similarly, zinc deprivation associated with malnutrition, ageing or disease induces the production of pro-inflammatory cytokines IL-1β and TNF-α involving both, redox-dependent mechanisms and chromatin rearrangements (Wessels, Haase et al. 2013). Still, more research is needed to determine whether zinc deficiency during early life is another factor that primes microglia for hyperactivation in later life.

Homocysteine (Hcy) is another metabolite regulated by diet and associated with neurodegeneration. Hcy is a byproduct of the one-carbon metabolism pathway which produces the methyl groups incorporated into DNA or proteins by conversion of SAM (S-adenosyl methionine) into SAH (S-adenosyl homocysteine). High levels of Hcy inhibit DNA-methyl transferase enzymes affecting the epigenome and the transcriptome and also impacting the responses to oxidation and inflammation (Ducker and Rabinowitz 2017). Increased Hcy levels have been associated with AD and PD progression (Christine, Auinger et al. 2018, Baroni, Bonetto et al. 2019) and importantly, can influence the transition from cognitive impairment into dementia (Reitz, Tang et al. 2009, Blasko, Hinterberger et al. 2012). Since the levels of Hcy in healthy subjects are primarily determined by dietary intake of methionine, folate and vitamin B12; early life nutritional stress may prime the organism for neurological disease (Moretti and Caruso 2019). Hcy induces proliferation and activation of microglia in vitro and increases oxidative stress and microgliosis in rodent models (Zou, Zhao et al. 2010). Furthermore, Li et al. showed that hyperhomocysteinemia induced in mice by supplementation with methionine, increases the level of inflammation markers TNF$\alpha$ and IL-1$\beta$ in plasma, along with changes in the expression of cystathionine $\gamma$-lyase (CSE) in macrophages. Importantly, in vitro testing showed that these alterations were mediated by epigenetic changes - namely increased DNA methylation at CSE promoter - and may serve as a form of macrophage memory (Li, Li et al. 2015). Hcy was implicated in chronic stress-induced A$\beta$ protein accumulation due to APP misprocessing and with cognitive decline in adult rats further supporting the role of Hcy in AD via solely dietary or stress-triggered signaling cascades (Xie, Zhao et al. 2016). More research is needed to understand the potential impact of imbalanced micronutrient intake and its interplay with stress exposures during early life on Hcy metabolism and whether this early insult will sustain long-lasting deleterious effects in the brain via microglial priming.

**7. Can early-life programming of microglia be reversed? Looking for new therapeutic targets.**

While an adverse prenatal environment seems to increase the risk for neurodegeneration in later life, exposure to an enriched environment (EE) during early life may reduce the odds of neurodegeneration. Furthermore, this mechanism is mediated by microglia. EE is known to induce neurogenesis (Clemenson, Lee et al. 2015), and some studies suggest that EE can impact the number of microglia in cortex and hippocampus, also affecting their inflammatory and phagocytic activity (Williamson, Chao et al. 2012). A recent study showed that EE can modulate microglia in the dentate gyrus reducing the inflammatory response of microglia after exposure to A$\beta$ oligomers (Xu, Gelyana et al. 2016). Not surprisingly, EE-associated behavior reduces chronic stress, a pathway we invoked repeatedly in this review as the likely common culprit of multi-hit etiology which manifests a disease susceptibility into a clinical neurodegenerative phenotype.

Early life malnutrition may have a deleterious impact on the brain by increasing inflammation, but its mechanism of action also opens another opportunity for intervention. Manipulation of dietary intake of folate and omega-3 seems to reduce the effects of ELS on memory and cognition. For instance, manipulation of the ratios of long-chain polyunsaturated fatty acids in the dietary intake serves to increase the availability of omega-3 and has protective effects against ELS-associated cognitive impairment via modulation of microglia in the hippocampus (Yam, Schipper et al. 2019).

Importantly, international consensus was recently reached in declaring elevated plasma levels of Hcy as a modifiable risk factor for dementia in the elderly. This factor has a population attributable risk that ranges from 4.1% to 30% and can be safely treated by supplementation with vitamin B (Smith, Refsum et al. 2018). Moreover, a recent study by Velazquez et al. reported beneficial effects of maternal choline supplementation in reducing Hcy levels in the offspring. Hcy reduction was associated with amelioration of AD pathology in transgenic mice models, including reduction of A$\beta$ aggregates and microgliosis and improvement in cognitive function. The most remarkable finding of this study is that the benefits of diet manipulation extended across generations, and animals that were never exposed to choline supplementation but were born from the offspring of treated parents, still benefited from the intervention and showed a milder AD phenotype (Velazquez, Ferreira et al. 2019). Taking in all, these results suggest that early life dietary interventions may potentially revert adverse environmental cues, a mechanism that may be associated with microglial memory and epigenetic remodeling.

Pioneering work by Ben Barres and Beth Stevens labs highlighted the role of the complement pathway in the microglial-neuronal interactions which make the brain more vulnerable to AD over time (Schafer, Lehrman et al. 2012, Hong, Beja-Glasser et al. 2016). Similar signaling pathways, particularly the iron homeostasis pathway, mediate microglial memory of many different *in utero* insults; even when the adverse event is clinically asymptomatic (Cortes, Cao et al. 2017). Hyperactive microglia may prevent synaptogenesis thus predisposing to AD in later life (Stephan, Barres et al. 2012, Hong, Beja-Glasser et al. 2016). In primary fetal ovine microglia cultures, agonistic stimulation of the α7nACh receptor up-regulated C3AR1 (Cortes, Cao et al. 2017). These findings warrant further investigation as pro-cholinergic drugs are used in the treatment of AD symptoms although the consequences of cholinergic stimulation on microglial signaling are not yet well understood. Will selective blocking rather than enhancing α7nAChR signaling in microglia be potentially more beneficial for mitigating the AD progression by reducing the pace of synaptic degradation?

## 8. Conclusions

We provide evidence supporting the role of early life programming in the susceptibility for adult neurodegenerative diseases while highlighting questions and proposing avenues for future research to advance our understanding of this fundamental process. The key elements of this phenomenon are chronic stress, neuroinflammation triggering microglial polarization, microglial memory and their connection to neurodegeneration.

While this review focused on microglia, a full understanding of brain programming predisposing to neurodegenerative diseases will require incorporation of astrocytes and oligodendroglia and their dynamic interactions with microglia and neuronal cells (Hong, Beja-Glasser et al. 2016, Liddelow, Guttenplan et al. 2017, Dutta, Woo et al. 2018).

We synthesize the reviewed mechanisms in Figure 1 which integrates the compounded impact of PS as well as the accompanying and independently occurring systemic and neuroinflammation on the glial energy reserves. The evolving metabolic phenotype of the stressed brain may represent the bottleneck substrate of the emergent, inherited and acquired resilience phenotype which, in turn, over multiple, subclinical hits with PS and/or inflammation modulates the progression of immunosenescence and the risk for neurodegeneration. Such framework offers a number of opportunities for early therapeutic interventions with long-term health benefits reducing the risk for neurodegeneration by modulating microglial immunometabolism.

**ACKNOWLEDGEMENTS**
This work was partially supported by grants NS092803 and NS104013 from NINDS to P.D. MGF is funded by Canadian Institutes of Health Research (CIHR). MCA is funded, in part, by Hans Fisher Senior Fellowship (Institute for Advanced Study-Technical University Munich (IAS-TUM), Munich, Germany).

**Figure legends**

**Figure 1.** Spatiotemporal integration of prenatal stress and the accompanying or independently occurring systemic- and neuro-inflammation on glial energy reserves that may modulate the risk for neurodegeneration via modulation of the pace or extent of immunosenescence.

*(1),* maternal-fetal circulation as an example of the interface for stress or inflammation/infection transfer of endogenous or exogenous biological mediators such as stress hormones, cytokines, bacteria or viruses (other interfaces are known for transfer of maternal exposures onto the fetus such as direct bacterial ascension into the fetal compartment or maternal-fetal heart rate entrainment which is likely biophysical). These factors induce or modulate changes in microglia in the fetal brain mediated by epigenetic changes *(2).*
*(3),* cogwheel representation of interlocking, spatiotemporally (across organs/systems and during fetal/child growth and adulthood) distributed effectors which together shape the individual phenotype of the adult brain
*(4),* over a sequence of repetitive, potentially asymptomatic modifications in (1) - (3); child and adult neuro- and glial development are shaped with variable degrees of susceptibility to neurodegeneration.

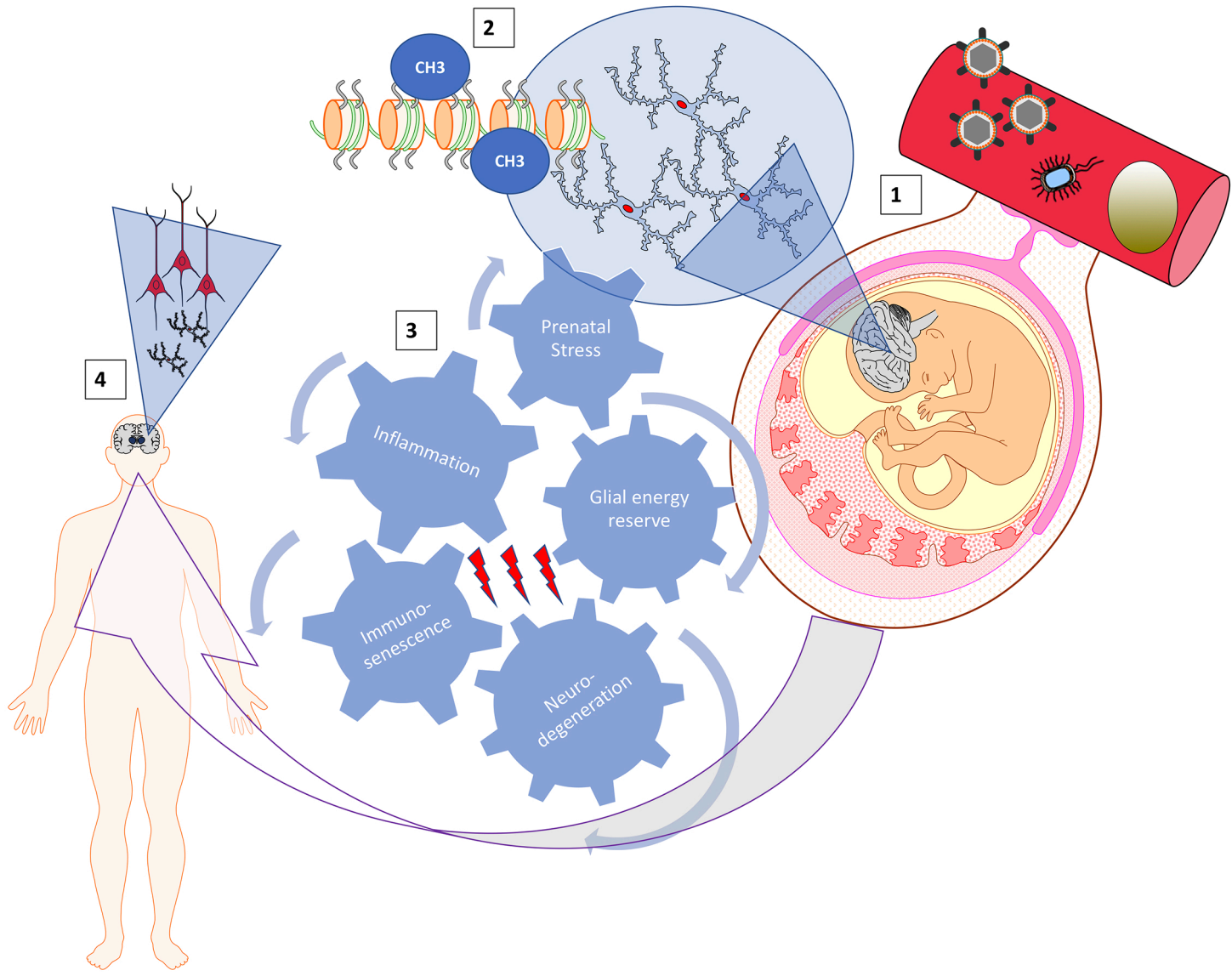

Munoz-Fernadez, R. Sussams, H. Lin, T. J. Fairchild, Y. A. Benito, C. Holmes, H. Karamujic-Comic, M. P. Frosch, H. Thonberg, W. Maier, G. Roschupkin, B. Ghetti, V. Giedraitis, A. Kawalia, S. Li, R. M. Huebinger, L. Kilander, S. Moebus, I. Hernandez, M. I. Kamboh, R. Brundin, J. Turton, Q. Yang, M. J. Katz, L. Concari, J. Lord, A. S. Beiser, C. D. Keene, S. Helisalmi, I. Kloszewska, W. A. Kukull, A. M. Koivisto, A. Lynch, L. Tarraga, E. B. Larson, A. Haapasalo, B. Lawlor, T. H. Mosley, R. B. Lipton, V. Solfrizzi, M. Gill, W. T. Longstreth, Jr., T. J. Montine, V. Frisardi, M. Diez-Fairen, F. Rivadeneira, R. C. Petersen, V. Deramecourt, I. Alvarez, F. Salani, A. Ciaramella, E. Boerwinkle, E. M. Reiman, N. Fievet, J. I. Rotter, J. S. Reisch, O. Hanon, C. Cupidi, A. G. Andre Uitterlinden, D. R. Royall, C. Dufouil, R. G. Maletta, I. de Rojas, M. Sano, A. Brice, R. Cecchetti, P. S. George-Hyslop, K. Ritchie, M. Tsolaki, D. W. Tsuang, B. Dubois, D. Craig, C. K. Wu, H. Soininen, D. Avramidou, R. L. Albin, L. Fratiglioni, A. Germanou, L. G. Apostolova, L. Keller, M. Koutroumani, S. E. Arnold, F. Panza, O. Gkatzima, S. Asthana, D. Hannequin, P. Whitehead, C. S. Atwood, P. Caffarra, H. Hampel, I. Quintela, A. Carracedo, L. Lannfelt, D. C. Rubinsztein, L. L. Barnes, F. Pasquier, L. Frolich, S. Barral, B. McGuinness, T. G. Beach, J. A. Johnston, J. T. Becker, P. Passmore, E. H. Bigio, J. M. Schott, T. D. Bird, J. D. Warren, B. F. Boeve, M. K. Lupton, J. D. Bowen, P. Proitsi, A. Boxer, J. F. Powell, J. R. Burke, J. S. K. Kauwe, J. M. Burns, M. Mancuso, J. D. Buxbaum, U. Bonuccelli, N. J. Cairns, A. McQuillin, C. Cao, G. Livingston, C. S. Carlson, N. J. Bass, C. M. Carlsson, J. Hardy, R. M. Carney, J. Bras, M. M. Carrasquillo, R. Guerreiro, M. Allen, H. C. Chui, E. Fisher, C. Masullo, E. A. Crocco, C. DeCarli, G. Bisceglio, M. Dick, L. Ma, R. Duara, N. R. Graff-Radford, D. A. Evans, A. Hodges, K. M. Faber, M. Scherer, K. B. Fallon, M. Riemenschneider, D. W. Fardo, R. Heun, M. R. Farlow, H. Kolsch, S. Ferris, M. Leber, T. M. Foroud, I. Heuser, D. R. Galasko, I. Giegling, M. Gearing, M. Hull, D. H. Geschwind, J. R. Gilbert, J. Morris, R. C. Green, K. Mayo, J. H. Growdon, T. Feulner, R. L. Hamilton, L. E. Harrell, D. Drichel, L. S. Honig, T. D. Cushion, M. J. Huentelman, P. Hollingworth, C. M. Hulette, B. T. Hyman, R. Marshall, G. P. Jarvik, A. Meggy, E. Abner, G. E. Menzies, L. W. Jin, G. Leonenko, L. M. Real, G. R. Jun, C. T. Baldwin, D. Grozeva, A. Karydas, G. Russo, J. A. Kaye, R. Kim, F. Jessen, N. W. Kowall, B. Vellas, J. H. Kramer, E. Vardy, F. M. LaFerla, K. H. Jockel, J. J. Lah, M. Dichgans, J. B. Leverenz, D. Mann, A. I. Levey, S. Pickering-Brown, A. P. Lieberman, N. Klopp, K. L. Lunetta, H. E. Wichmann, C. G. Lyketsos, K. Morgan, D. C. Marson, K. Brown, F. Martiniuk, C. Medway, D. C. Mash, M. M. Nothen, E. Masliah, N. M. Hooper, W. C. McCormick, A. Daniele, S. M. McCurry, A. Bayer, A. N. McDavid, J. Gallacher, A. C. McKee, H. van den Bussche, M. Mesulam, C. Brayne, B. L. Miller, S. Riedel-Heller, C. A. Miller, J. W. Miller, A. Al-Chalabi, J. C. Morris, C. E. Shaw, A. J. Myers, J. Wiltfang, S. O'Bryant, J. M. Olichney, V. Alvarez, J. E. Parisi, A. B. Singleton, H. L. Paulson, J. Collinge, W. R. Perry, S. Mead, E. Peskind, D. H. Cribbs, M. Rossor, A. Pierce, N. S. Ryan, W. W. Poon, B. Nacmias, H. Potter, S. Sorbi, J. F. Quinn, E. Sacchinelli, A. Raj, G. Spalletta, M. Raskind, C. Caltagirone, P. Bossu, M. D. Orfei, B. Reisberg, R. Clarke, C. Reitz, A. D. Smith, J. M. Ringman, D. Warden, E. D. Roberson, G. Wilcock, E. Rogaeva, A. C. Bruni, H. J. Rosen, M. Gallo, R. N. Rosenberg, Y. Ben-Shlomo, M. A. Sager, P. Mecocci, A. J. Saykin, P. Pastor, M. L. Cuccaro, J. M. Vance, J. A. Schneider, L. S. Schneider, S. Slifer, W. W. Seeley, A. G. Smith, J. A. Sonnen, S. Spina, R. A. Stern, R. H. Swerdlow, M. Tang, R. E. Tanzi, J. Q. Trojanowski, J. C. Troncoso, V. M. Van Deerlin, L. J. Van Eldik, H. V. Vinters, J. P.